\documentclass[
    aps,
    prd,
    nofootinbib,
    superscriptaddress,
    10pt,
    notitlepage,
    preprintnumbers,
    twocolumn]{revtex4-1}

\usepackage{anyfontsize}
\usepackage{amssymb}
\usepackage{amsmath}
\usepackage{amsfonts}
\usepackage{amsbsy}

\usepackage{newtxtext}
\usepackage{newtxmath}

\usepackage{tensor}
\usepackage{mathrsfs}
\usepackage{bm}

\usepackage[utf8]{inputenc}

\usepackage{graphicx}
\usepackage{epsfig}
\usepackage{epstopdf}

\usepackage[linktocpage,breaklinks]{hyperref}
\usepackage[usenames,dvipsnames]{xcolor}

\definecolor{romared}{RGB}{142,0,28}
\hypersetup{colorlinks=true,
            citecolor=romared,
            linkcolor=romared,
            urlcolor=romared}

\newcommand{\be}{\begin{equation}}
\newcommand{\ee}{\end{equation}}

\def\be{\begin{equation}}
\def\ee{\end{equation}}
\newcommand{\beq}{\begin{eqnarray}}
\newcommand{\eeq}{\end{eqnarray}}

\begin{document}

\title{Parametrized black hole quasinormal ringdown. \\II. Coupled equations and quadratic corrections for nonrotating black holes}

\author{Ryan McManus}
\email{rmcmanu3@jhu.edu}
\affiliation{Department of Physics and Astronomy, Johns Hopkins University, 3400 N. Charles Street, Baltimore, MD 21218, US}

\author{Emanuele Berti}
\email{berti@jhu.edu}
\affiliation{Department of Physics and Astronomy, Johns Hopkins University, 3400 N. Charles Street, Baltimore, MD 21218, US}

\author{Caio F. B. Macedo}
\email{caiomacedo@ufpa.br}
\affiliation{Campus Salin\'opolis, Universidade Federal do Par\'a, Salin\'opolis, Par\'a, 68721-000 Brazil}

\author{Masashi Kimura}
\email{mkimura@rikkyo.ac.jp}
\affiliation{Department of Physics, Rikkyo University, Tokyo 171-8501, Japan}

\author{Andrea Maselli}
\email{andrea.maselli@roma1.infn.it}
\affiliation{Dipartimento di Fisica, ``Sapienza'' Universit\`a di Roma \& Sezione INFN Roma1, P.A. Moro 5, 00185, Roma, Italy}

\author{Vitor Cardoso}
\email{vitor.cardoso@ist.utl.pt}
\affiliation{CENTRA, Departamento de F\'{\i}sica, Instituto Superior T\'ecnico -- IST, Universidade de Lisboa -- UL,
Avenida Rovisco Pais 1, 1049 Lisboa, Portugal}
\affiliation{Theoretical Physics Department, CERN 1 Esplanade des Particules, Geneva 23, CH-1211, Switzerland}

\begin{abstract}
Linear perturbations of spherically symmetric spacetimes in general relativity are described by radial wave equations, with potentials that depend on the spin of the perturbing field. In previous work~\cite{Cardoso:2019mqo} we studied the quasinormal mode spectrum of spacetimes for which the radial potentials are slightly modified from their general relativistic form, writing generic small modifications as a power-series expansion in the radial coordinate. We assumed that the perturbations in the quasinormal frequencies are linear in some perturbative parameter, and that there is no coupling between the perturbation equations. In general, matter fields and modifications to the gravitational field equations lead to coupled wave equations. Here we extend our previous analysis in two important ways: we study second-order corrections in the perturbative parameter, and we address the more complex (and realistic) case of coupled wave equations. We highlight the special nature of coupling-induced corrections when two of the wave equations have degenerate spectra, and we provide a ready-to-use recipe to compute quasinormal modes. We illustrate the power of our parametrization by applying it to various examples, including dynamical Chern-Simons gravity, Horndeski gravity and an effective field theory-inspired model.
\end{abstract}

\preprint{RUP-19-17}

\date{\today}
\maketitle

\section{Introduction}\label{sec:intro}

There are experimental and conceptual reasons to expect that general relativity (GR) and the standard model of particle physics should be modified at some level. Most modifications of GR involve additional gravitational degrees of freedom and higher-order curvature corrections~\cite{Berti:2015itd,Barack:2018yly}.  Black holes (BHs) are a promising experimental playground to reveal or constrain these modifications. For example, light bosonic fields can trigger nonperturbative effects in astrophysical BHs via superradiance, affecting the spin distribution of astrophysical BHs and leading to potentially observable gravitational-wave and electromagnetic signatures~\cite{Arvanitaki:2010sy,Pani:2012vp,Brito:2017wnc,Brito:2017zvb,Baumann:2018vus,Hannuksela:2018izj,Ikeda:2019fvj,Berti:2019wnn}.

In the absence of large, smoking-gun effects, we must rely on precision measurements. Astrophysical BHs in GR are remarkably simple, being characterized only by their mass and spin by virtue of the so-called ``no-hair theorems''~\cite{Bekenstein:1972ny,Bekenstein:1995un,Sotiriou:2011dz,Hui:2012qt,Herdeiro:2015waa,Cardoso:2016ryw}. As such, they are ideal laboratories for precision measurements: any deviation from this simplicity is a potential hint of new physics. In particular, the relaxation of BH spacetimes to their equilibrium configuration in GR is very simple.  Consider for example two BHs merging to form a single spinning BH, a process of particular relevance for gravitational-wave astronomy.  The merger can be highly dynamical and violent. However, according to GR, at late times the remnant must be a slightly perturbed Kerr solution, described by only two parameters: its mass and spin.  The relaxation to a Kerr remnant is well described by linear perturbation theory. This is known as the ``ringdown'' stage, where the gravitational-wave amplitude consists of a superposition of exponentially damped sinusoids or ``quasinormal modes'' (QNMs) with characteristic frequencies and damping times~\cite{Kokkotas:1999bd,Berti:2009kk}.  A superposition of QNMs accurately describes the merger waveform even before the peak of the gravitational-wave emission~\cite{Leaver:1986gd,Andersson:1995zk,Berti:2006wq,Zhang:2013ksa,Baibhav:2017jhs,Brito:2018rfr,Giesler:2019uxc,Isi:2019aib}.  Therefore, one possible test of GR consists of testing the consistency between vacuum, linearized GR predictions and the observed QNM spectrum. This idea is now commonly called ``black hole spectroscopy''~\cite{Detweiler:1980gk,Dreyer:2003bv,Berti:2005ys,Berti:2007zu}.

The general procedure to test a given modification of GR is to find BH solutions, compute their QNM spectrum, and finally constrain deviations of the spectrum from the GR predictions through gravitational-wave observations.  The differential equations that must be solved to determine the QNM spectrum have a relatively simple, ``universal'' structure. A general parametrization of Schwarzschild perturbations induced by scalar, vector and tensor fields shows that linearized perturbations always lead to wave-like equations~\cite{Tattersall:2017erk,Tattersall:2018map,Tattersall:2018nve,Tattersall:2019pvx}. However, in general these wave-like equations are coupled. Verifying which theories lead to coupled perturbation equations is a laborious task, but some known examples in the literature include the low-energy limit of string-motivated theories, such as Einstein-dilaton-Gauss-Bonnet~\cite{Garfinkle:1990qj,Mignemi:1992nt,Pani:2009wy,Blazquez-Salcedo:2016enn} and dynamical Chern-Simons (dCS) gravity~\cite{Jackiw:2003pm,Alexander:2009tp,Cardoso:2009pk,Pani:2013pma,Kimura:2018nxk}.  Coupling also occurs in effective field theory (EFT) modifications of GR~\cite{Endlich:2017tqa,Cardoso:2018ptl,Kuntz:2019zef,Franciolini:2018uyq}.

Even if GR is the correct theory of gravity, matter fields can couple with each other, and the perturbations of these fields will in general be coupled. This happens, for instance, in the Einstein-Maxwell system~\cite{Leaver:1990zz,Berti:2005eb,Pani:2013ija,Pani:2013wsa,Mark:2014aja,Zimmerman:2014aha,Dias:2015wqa} or for axionic fields in charged BH backgrounds~\cite{Olive:2007aj,Stadnik:2015kpa,Boskovic:2018lkj,Ikeda:2019fvj}.

We have recently computed QNM frequencies for scalar, vector and tensor perturbations of a spherically symmetric spacetime which can be described as small deviations from the corresponding GR perturbation equations (see Ref.~\cite{Cardoso:2019mqo}, henceforth Paper I).  We wrote deviations in the corresponding radial potentials as a power series in the (inverse) radial coordinate. We found that corrections to the QNM frequencies are (to leading order) linear in these perturbations, and we computed the coefficients that determine these corrections. Here we extend these results by calculating {\em quadratic} corrections in the perturbative potentials, as well as the corrections that arise from {\em coupling} power-law perturbative corrections between the scalar, vector, polar (even-parity) and axial (odd-parity) gravitational perturbation equations in GR.  We still work under the assumption that the background solution is nonspinning (although, as shown in Paper I, the formalism can be applied to spinning black holes in the slow-rotation limit~\cite{Pani:2012bp,Pani:2013pma}) and that the perturbation equations are separable.

\subsection{Executive summary}

Our starting point is a generalized, matrix-valued master equation for the coupled radial perturbations induced by $N$ fields ${\bm\Phi}=\{\Phi_i\}$ ($i=1,\,\dots,\,N$):
\begin{equation}
    \label{eq:waveEQ}
    f \frac{d}{dr}\left(f \frac{d {\bm\Phi}}{dr}\right) +[\omega^2 - f \mathbf{V}] {\bm\Phi} = 0\,.
\end{equation}
Here $f = 1 - r_H / r$, $r_H=2M$ is the horizon radius, $\omega$ is the complex QNM frequency, and $\mathbf{V}(r)=V_{ij}(r)$ is a $N\times N$ matrix of radial potentials.\footnote{In principle the coupled perturbation equations may also involve ``friction-like'' terms of the form $f\mathbf{Z}\partial_r {\bm \Phi}$, i.e. terms of first order in radial derivatives~\cite{Tattersall:2018nve}. However the matrix $\mathbf{Z}$ can be reabsorbed into $\mathbf{V}$ through field redefinitions, as shown in Appendix~\ref{app:NoFriction}.} The factor of $f$ ensures that the effective potential terms vanish at the horizon.
We assume the background spacetime to be asymptotically flat, and we parametrize $\mathbf{V}$ 
as a sum of the GR potentials $V_{ij}^{\rm GR}$ and small power-law series corrections $\delta V_{ij}$: 
\beq
V_{ij} &=& V_{ij}^{\rm GR} + \delta V_{ij} \,, \\
\delta V_{ij} &=& \frac{1}{r_H^2} \sum^{\infty}_{k=0} \alpha_{ij}^{(k)} \left( \frac{r_H}{r} \right)^k \label{eq:dV}\,. \label{eq:PotentialMatrix}
\eeq
Here $V_{ij}^{\rm GR}$ denotes the potentials describing massless spin
$s=0,\,1,\,2$ perturbations in GR. Usually, $V_{ij}^{\rm GR}=0$ for
$i\neq j$ and $V_{ii}^{\rm GR} \neq 0$, since the fields decouple in
GR (but see~\cite{Leaver:1990zz,Boskovic:2018lkj} for
counterexamples). The coefficients $\alpha^{(k)}_{ij}$ are independent
of $r$, but they may be functions of
$\omega$~\cite{Cardoso:2009pk,Cardoso:2019mqo}. For $k\ge 1$ the
potential matrix vanishes at spatial infinity, i.e. $V_{ij}(r) \to 0 $
as $r\to\infty$. Perturbations with $k=0$ tend to a constant value,
$\delta V_{ij}\to \alpha_{ij}^{(0)}/r_H^2$. For simplicity, we neglect
off-diagonal contributions that fall off slower than the GR potentials
[cf. Eqs.~\eqref{eq:EffVm}--\eqref{eq:EffVs} below]:
$\alpha_{ij}^{(0)}=\alpha_{ij}^{(1)}=0$ for $i\neq j$.

The terms $\delta V_{ij}$ in the master equation \eqref{eq:waveEQ} will, in general, modify the GR QNM frequencies $\omega_0$ 
by a correction that is perturbatively small in $\alpha^{(k)}_{ij}$.  The key result of this work is that the corrected QNM frequencies 
read
\begin{equation}
\label{eq:OmegaExpansion}
    \omega \approx \omega_0 +  \alpha_{ij}^{(k)} d^{ij}_{(k)} + \alpha_{ij}^{(k)} \alpha'{}_{pq}^{(s)} d^{ij}_{(k)}d^{pq}_{(s)} 
    + \frac{1}{2} \alpha_{ij}^{(k)} \alpha_{pq}^{(s)} e^{ijpq}_{(ks)} \,,
\end{equation}
where $e^{ijpq}_{(ks)}=e^{pqij}_{(sk)}$,  $i,j,p,q=1,\ldots,N$, $k,s = 0,\ldots,\infty$, and we use the Einstein summation convention.  A prime denotes a derivative with respect to $\omega$, with all $\alpha_{pq}^{(s)}$ and $\alpha'{}_{pq}^{(s)}$ evaluated at $\omega_0$. The derivation of Eq.~\eqref{eq:OmegaExpansion} is presented in Appendix~\ref{app:omegaExpansion}.

The values of $\omega_0$ for the tensor, vector and scalar perturbations~\cite{Berti:2005ys,Berti:2009kk} and the coefficients $d^{ii}_{(k)}$~\cite{Cardoso:2019mqo} are available online~\cite{GRITJHU}.  The values of $d^{ij}_{(k)}$ and $e^{ijpq}_{(ks)}$ for these same perturbations were first computed in this work, and they are also available online~\cite{GRITJHU}. 
We stress that $d^{ij}_{(k)}$ and $e^{ijpq}_{(ks)}$ depend on the unperturbed potentials $V_{ij}^{\rm GR}$ and on the unperturbed QNM frequency $\omega_0$.

Equation~\eqref{eq:OmegaExpansion} allows for the efficient calculation of the QNM frequencies to quadratic order through simple multiplications and additions.  While this expression may look complex, its use is trivial: the prefactors $\alpha_{pq}^{(s)}$ and their derivatives are, in principle, all independent, with their values prescribed by the specific theory in question (cf. Section~\ref{sec:Examples} for examples).

\subsection{Plan of the paper}\label{sec:plan}

The plan of the paper is as follows.
In Section~\ref{sec:Background} we present the eigenvalue problem for the QNM frequencies, and we briefly review the numerical method to find them.
In Section~\ref{sec:QuadraticUncoupled} we compute quadratic corrections for uncoupled fields.
In Section~\ref{sec:NonDegCoupled} we show that coupling fields whose spectra are nondegenerate leads to quadratic corrections in the QNM frequencies.
In Section~\ref{sec:DegCoupled} we show that coupling fields whose spectra are degenerate leads to corrections in the QNM frequencies which are linear in the perturbation parameter.
Finally, in Section~\ref{sec:Examples} we apply the formalism to some specific examples: dCS gravity~\cite{Cardoso:2009pk,Kimura:2018nxk}, Horndeski gravity~\cite{Tattersall:2018nve}, and an EFT-inspired model~\cite{Cardoso:2018ptl}.
In Section~\ref{sec:Conclusion} we discuss some limitations of our analysis and directions for future work.

To improve readability, we relegate several technical results to the appendices.
As already mentioned, in Appendix \ref{app:NoFriction} we demonstrate that friction-like terms (containing first derivatives of the fields) can be reabsorbed in the definition of the potentials, and in Appendix \ref{app:omegaExpansion} we derive Eq.~(\ref{eq:OmegaExpansion}).
In Appendix \ref{app:threeFields} we look at the case of three fields. There we show that (i) QNM frequency corrections arising from the coupling of two fields are independent of the total number of coupled fields up to quadratic order, and (ii) when two uncoupled fields have nondegenerate spectra, a linear coupling between the fields gives rise to quadratic QNM frequency corrections.
Finally, in Appendix~\ref{app:Degenerateomega1derivation} we prove that when two uncoupled fields have degenerate spectra, a linear coupling between the fields gives rise to linear QNM frequency corrections.

\section{Background}\label{sec:Background}
\subsection{Quasinormal modes in general relativity}\label{sec:Motivation}

Gravitational perturbations of the Schwarzschild geometry in GR can be classified by their behavior under parity.\footnote{It is possible to construct definite parity perturbations even in the Kerr background: see e.g. Appendix C of~\cite{Nichols:2012jn}.} It is common to classify the metric perturbations as odd (or axial, or Regge-Wheeler) and even (or polar, or Zerilli).
These are governed by master variables $\Phi_\pm$ which obey the master equations~\citep{Regge:1957td,Zerilli:1970se}
\begin{equation}
f \frac{d}{dr}\left(f \frac{d\Phi_\pm}{dr}\right) +[\omega^2 - f V_{\pm}] \Phi_\pm = 0\,.\label{eq:EvenOddEq}
\end{equation}
The effective potential for odd perturbations reads 
\begin{equation}
\label{eq:EffVm}
V_- = \frac{\ell (\ell +1)}{r^2} - \frac{3 r_H}{r^3}\,,
\end{equation}
while the effective potential for even perturbations is
\begin{equation}
\label{eq:EffVp}
V_+ = \frac{9 \lambda r_H^2 r + 3 \lambda^2 r_H r^2 + \lambda^2 (\lambda+2) r^3 + 9 r_H^3}{r^3(\lambda r + 3 r_H)^2}\,,
\end{equation}
where $\lambda = \ell^2+\ell-2$, and $\ell$ is an angular harmonic index labeling the tensorial spherical harmonics used to separate the angular dependence of the perturbations. 
These two potentials are, quite remarkably, isospectral~\cite{Chandrasekhar:1975zza,Chandrasekhar:1985kt}, and maintaining isospectrality under generic perturbations of the potentials requires fine tuning~\cite{Cardoso:2019mqo}.

Perturbations of a Schwarzschild background induced by scalar and vector fields are of interest not only in modified gravity theories (that in general introduce additional degrees of freedom), but also in phenomena that involve coupling between gravitational and nongravitational fields.  In GR, Schwarzschild perturbations induced by fields $\phi_s$ of spin $s=0,\,1$ are also described by master equations similar to Eq.~\eqref{eq:EvenOddEq}, with potentials~\cite{Berti:2009kk}
\begin{equation}
\label{eq:EffVs}
V_s = \frac{\ell ( \ell + 1)}{r^2} + (1 - s^2) \frac{r_H}{r^3}\,.
\end{equation}
Note that $V_2=V_-$, i.e. the $s=2$ potential in Eq.~(\ref{eq:EffVs}) corresponds to odd gravitational perturbations.

\subsection{Calculation of the quasinormal frequencies}
\label{sec:numerics}
There are many techniques to compute QNM frequencies~\citep{Berti:2009kk,Pani:2013pma,Macedo:2016wgh}. Since we are striving for generality, here we follow a direct integration approach~\citep{Pani:2013pma}. The idea is to integrate the radial wave equations from the horizon to infinity given an initial guess of the QNM frequency, and to vary the frequency until Eq.~\eqref{eq:waveEQ} and the relevant boundary conditions are satisfied.  The values of the fields at the horizon must be specified to perform the integration.  The horizon fields form an $N$-dimensional vector ${\bm \Phi}_H^{(i)}$ and we can perform the integration for any basis of $N$ such ${\bm \Phi}_H^{(i)}=\left\{{\bm \Phi}^{(i)}_j\right\}$, with the final result independent of the choice of basis.  For simplicity, in our integrations we consider a basis such that ${\bm \Phi}^{(i)}_j = \delta^i_j$, and we then construct an $N \times N$ matrix $\mathbf{S}$ from the integration of the $N$ fields under these $N$ initial conditions.
The eigenvalues of Eq.~\eqref{eq:waveEQ} are then the complex roots of
\begin{equation}
\label{eq:detS}
{\cal S}(\omega)\equiv \det\mathbf{S}(\omega) = 0\,,
\end{equation}
which can be found numerically.

As we show below, by expanding Eq.~\eqref{eq:detS} with respect to the small parameters $\alpha_{ij}^{(k)}$ we get semianalytic expressions for the coefficients $d^{ij}_{(k)}$ and $e^{ijpq}_{(ks)}$ appearing in Eq.~\eqref{eq:OmegaExpansion}.

\begin{figure*}[thbp]
\includegraphics[width=\columnwidth]{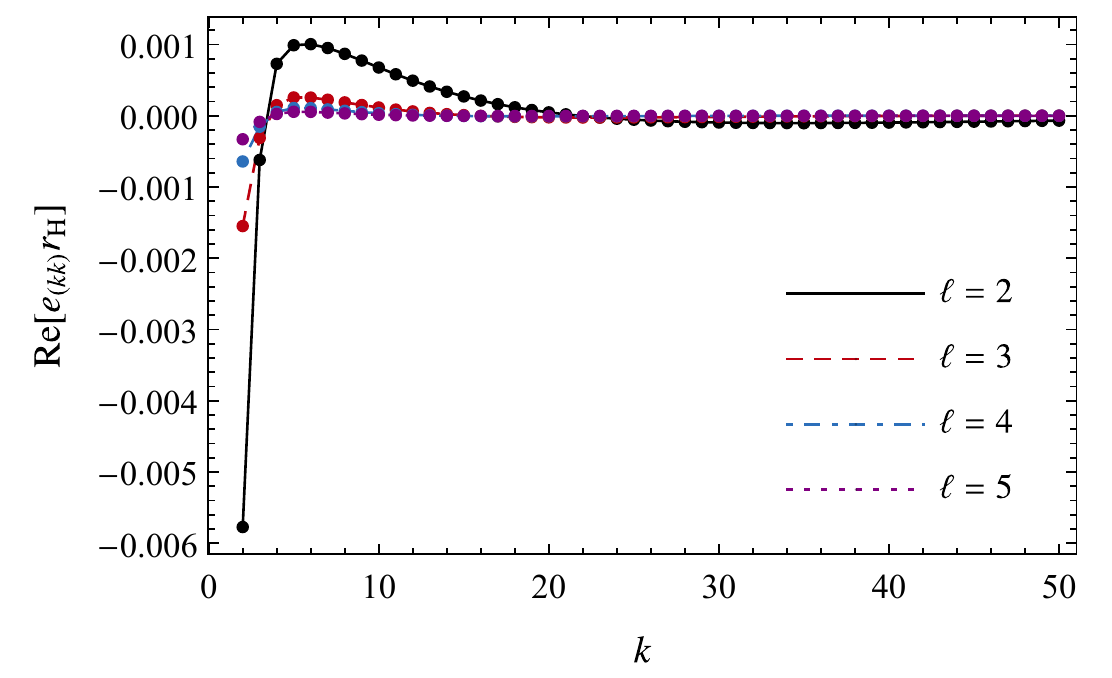}
\includegraphics[width=\columnwidth]{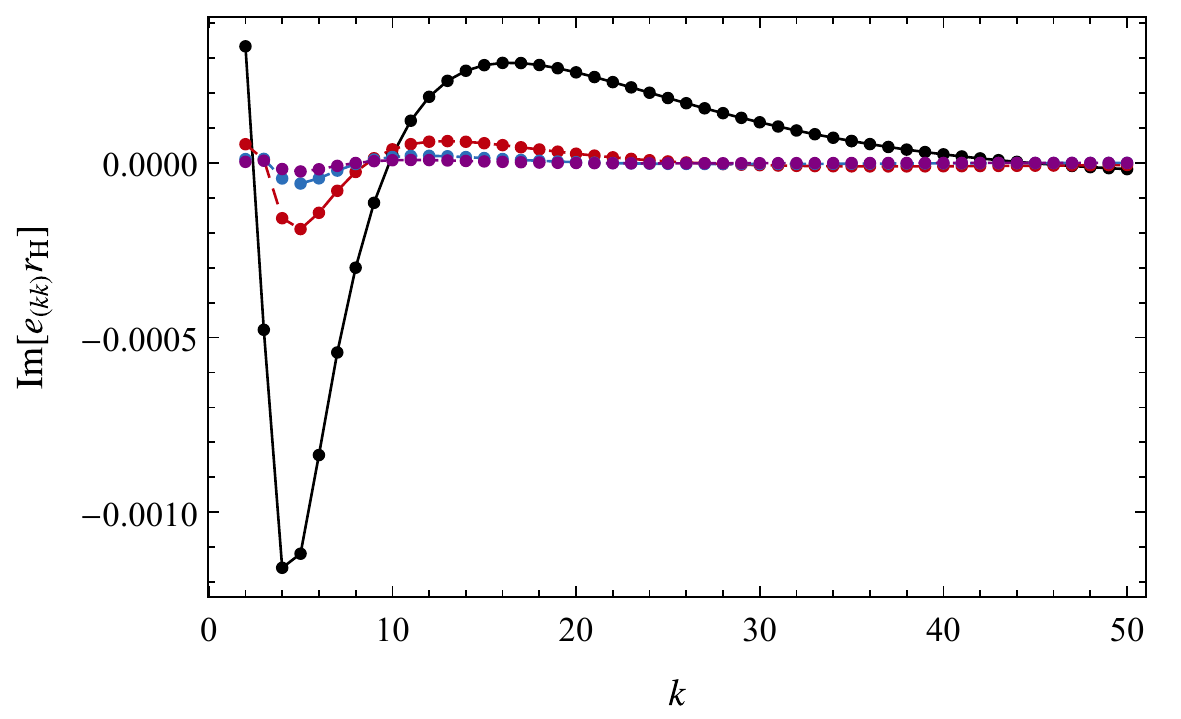}\\
\caption{Real and imaginary parts of $e_{(kk)}$ [Eq.~\eqref{eq:OmegaExpansion}] for axial gravitational perturbations with selected values of $\ell$ and $2\leq k\leq50$. 
}
\label{fig:ekkSingleField}
\end{figure*}
\begin{figure*}[thbp]
\includegraphics[width=\columnwidth]{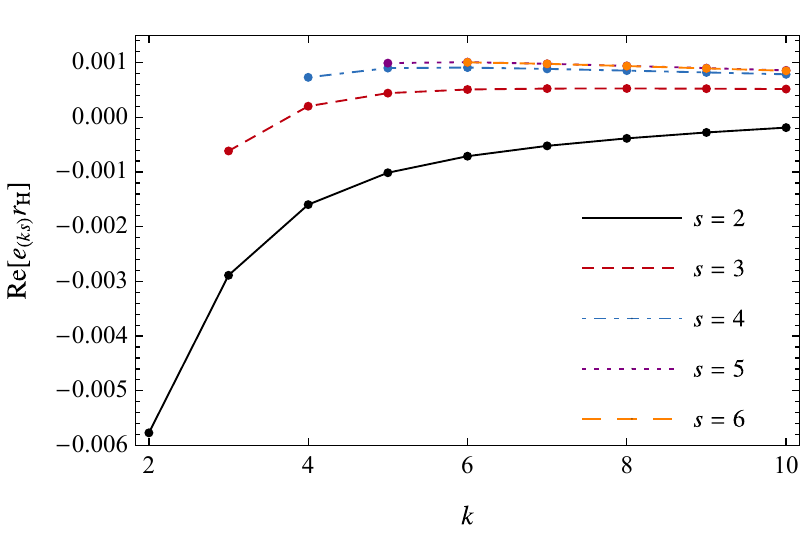}
\includegraphics[width=\columnwidth]{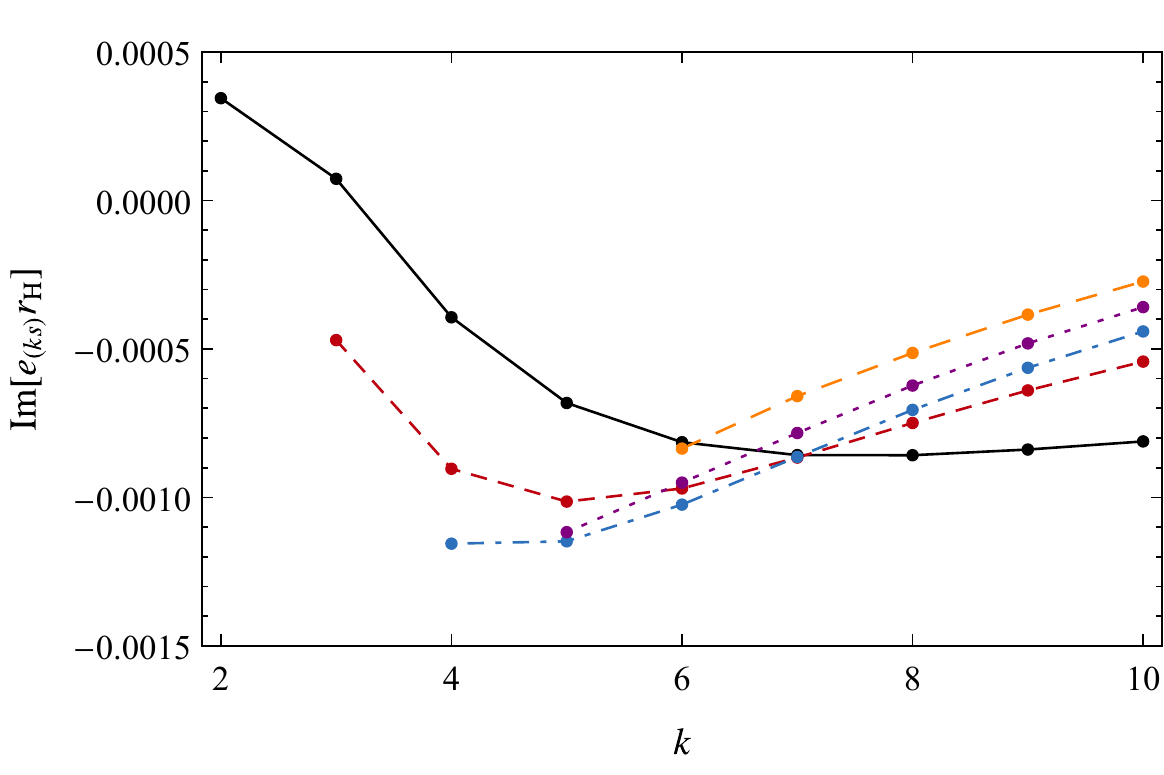}\\
\caption{
Real and imaginary parts of $e_{(ks)}$ [Eq.~\eqref{eq:OmegaExpansion}] for axial gravitational perturbations with $\ell=2$, showing the dependence on $k$ at fixed $s$. The leftmost value for each $s$ corresponds to $e_{(ss)}$, and it is bounded by the $\ell=2$ curves in Fig.~\ref{fig:ekkSingleField}.}
\label{fig:eksSingleField}
\end{figure*}

\section{Decoupled fields: quadratic corrections}
\label{sec:QuadraticUncoupled}

Let us start for simplicity from the case of a single field, and therefore a single nonzero $\alpha_{ij}^{(k)}=\alpha$ in the expansion \eqref{eq:PotentialMatrix}.  The QNM frequencies are the roots of Eq.~\eqref{eq:detS}, where the matrix $\mathbf{S}$ (now a scalar) is a function of both $\omega$ and $\alpha$.  As $\alpha$ and $\alpha'\equiv \partial_\omega \alpha$ are in principle independent, we can vary $\alpha$ while holding $\alpha'$ constant.
By expanding ${\cal S}$ up to second order in $\alpha$ we get
\begin{align}
    \label{eq:detSexpansion}
    {\cal S}(\omega,\alpha) =\left.{\cal S}\right|_{\alpha=0} +\alpha\left.\frac{d {\cal S}}{d\alpha}\right|_{\alpha=0} %
    +\frac{\alpha^2}{2}\left.\frac{d^2 {\cal S}}{d\alpha^2}\right|_{\alpha=0}\,.
\end{align}

Let us restrict the expansion of ${\cal S}(\omega,\alpha)$ to those points such that Eq.~\eqref{eq:detS} is satisfied. These points will describe a curve $\omega(\alpha)$ starting from $\omega=\omega_0$, $\alpha=0$.
Replacing the total derivative with respect to $\alpha$ with the directional derivative along $\omega(\alpha)$, one finds
\begin{align}
    0&={\cal S}(\omega_0,0)
        +\alpha\left(\frac{\partial}{\partial\alpha}
        +\frac{\partial\omega}{\partial\alpha}\frac{\partial}{\partial\omega}\right){\cal S}\bigg\vert_{\alpha=0}
        \nonumber \\
    &+\frac{\alpha^2}{2}\bigg(\frac{\partial}{\partial\alpha} +\frac{\partial\omega}{\partial\alpha}\frac{\partial}{\partial\omega}\bigg)^2{\cal S}\bigg\vert_{\alpha=0} \,.
    \label{eq:detSExpansionDirectional}
\end{align}
Each term in the expansion must vanish identically along the curve $\omega(\alpha)$.
Near $\alpha=0$, this curve is approximated by the expansion $\omega \approx \omega_0 + \alpha d + \alpha^2 e$: cf. Eq.~\eqref{eq:OmegaExpansion}.
By inserting this expansion into the linear term of Eq.~\eqref{eq:detSExpansionDirectional}, we find
\begin{equation}
    \left(\frac{\partial}{\partial\alpha}
        +(d+\alpha e) \frac{\partial}{\partial\omega}\right){\cal S}\bigg\vert_{\alpha=0}=0\,.
\end{equation}
We can now evaluate this expression at $(\omega_0,0)$ and solve for $d$:
\begin{equation}
  d = -
  \frac{1}{\left.\partial_\omega {\cal S}\right|_{(\omega_0,0)}}
  \left.\frac{\partial {\cal S}}{\partial\alpha}\right|_{(\omega_0,0)}
  \,.
    \label{eq:omega1Uncoupled}
\end{equation}
Following the same steps for the quadratic term in Eq.~\eqref{eq:detSExpansionDirectional} we find an expression for $e$:
\begin{equation}
    \label{eq:kkOmega2Uncoupled}
    e = -\frac{1} {\left. \partial_\omega {\cal S}\right|_{(\omega_0,0)}}
    \left.\left(\frac{\partial^2}{\partial\alpha^2}+2d\frac{\partial^2}{\partial\alpha\partial\omega} + d^2\frac{\partial^2}{\partial\omega^2}\right){\cal S}\right|_{(\omega_0,0)}\,.
\end{equation}
By construction, both of these expressions depend on the structure of ${\cal S}$ at $\alpha=0$, but not on the value of $\alpha$ in the expansion \eqref{eq:OmegaExpansion}, as long as $\alpha$ is small.
Therefore these results encode deviations from the GR spectrum in a theory-independent manner.

We now reinsert the $k$ index labeling specific power-law corrections to the potentials. For linear corrections we get
\begin{equation}
  d_{(k)} = -
  \frac{1}{\left.\partial_\omega {\cal S}\right|_{(\omega_0,0)}}
  \left.\frac{\partial {\cal S}}{\partial\alpha^{(k)}}\right|_{(\omega_0,0)}
  \,,
    \label{eq:dk}
\end{equation}
while quadratic corrections yield
\begin{align}
\label{eq:ksOmega2Uncoupled}
e_{(ks)} &=  -\frac{1}{\left.\partial_\omega {\cal S}\right|_{(\omega_0,0)}}
  \bigg(\frac{\partial^2}{\partial\alpha^{(k)}\partial\alpha^{(s)}}+d_{(k)}\frac{\partial^2}{\partial\alpha^{(s)}\partial\omega}
           \nonumber \\
         &+d_{(s)}\frac{\partial^2}{\partial\alpha^{(k)}\partial\omega} 
           +d_{(k)}d_{(s)}\frac{\partial^2}{\partial\omega^2}
    \bigg){\cal S}\mid_{(\omega_0,0)}
    \,.
\end{align}
Note that Eq.~\eqref{eq:kkOmega2Uncoupled} and Eq.~\eqref{eq:ksOmega2Uncoupled} agree when $k=s$, as  a result of the definition of $e_{(ks)}$ in the expansion \eqref{eq:OmegaExpansion}.

\begingroup
\setlength{\tabcolsep}{8pt} %
\begin{table}[]
\begin{tabular}{ccc}
\hline
\hline
    & Axial                 & Scalar \\
$k$ & $r_H\,e_{(kk)}$            & $r_H\,e_{(kk)} $ \\ 
\hline
\hline
2 & - 0.00580 + 0.000345 i & - 0.00303 - 0.0000263 i \\
3 & - 0.000620 - 0.000470 i & - 0.000964 0.000211 i \\
4 & 0.000731 - 0.00116 i & - 0.0000341 - 0.000162 i \\
5 & 0.000991 - 0.00112 i & 0.000250 - 0.000351 i \\
10 & 0.000678 + 0.0000227 i & 0.000263 - 0.0000671 i \\
\hline
\hline
\end{tabular}
\caption{
The quadratic frequency coefficients $e_{(kk)}$ for the {\it decoupled} odd-parity gravitational and scalar field perturbations with $\ell=2$, as defined in \eqref{eq:OmegaExpansion}.
The values for $e_{(ks)}$ with $k,s=0,\ldots,10$ and $\ell\leq5$ for scalar, vector, axial gravitational and polar gravitational perturbations are available online~\cite{GRITJHU}.}
\label{tab:quadraticUncoupled}
\end{table}
\endgroup

The linear corrections $d_{(k)}$ where found to five significant figures in Paper I. Here we present numerical results for the coefficients $e_{(ks)}$ for a single field perturbed by a power law potential.

In Table~\ref{tab:quadraticUncoupled} we list $r_H e_{(kk)}$ for axial and scalar gravitational perturbations with $\ell=2$ and selected values of $k$. The values for $\ell = 2,\ldots,5$ and $k,s=0,\ldots,10$ for the scalar, vector, axial gravitational and polar gravitational cases are available online~\cite{GRITJHU}.
The large-$k$ behavior of $e_{(kk)}$ for axial perturbations is shown in Fig.~\ref{fig:ekkSingleField}.
In Paper I we found that the linear coefficients $d_{(k)}$ in the large-$k$ limit are well approximated by 
\begin{equation}
    d_{(j)} \approx \frac{\kappa}{j^\beta}\sin(\gamma \ln j + \zeta)
\end{equation}
where $(\kappa,\beta,\gamma,\zeta)$ are numerical coefficients.
Assuming the same functional form for $e_{(kk)}$ for axial perturbations with $\ell=2$, the best-fit parameters are $\beta\approx1.7$ and $\gamma\approx2.4$, to be compared with $\beta\approx0.66$ and $\gamma\approx1.5$ for $d_{(k)}$.
At quadratic order, cross-term corrections $e_{(ks)}$ also contribute. 
Representative correction coefficients $e_{(ks)}$ for axial gravitational perturbations with $k\geq s$ and $\ell = 2$ are shown in Fig.~\ref{fig:eksSingleField}. 

These quadratic corrections are necessary when $\alpha>10^{-5}$. The real and imaginary parts of $e_{(kk)}$ and $d_{(k)}$ are of the same order of magnitude (compare Table I of Paper I with Table~\ref{tab:quadraticUncoupled} in this paper). If $\alpha^{(kk)} e_{(kk)}<10^{-5}d_{(k)}$ the quadratic correction would be smaller than the numerical error in the leading term, which is currently available to five significant figures.  By this argument, we expect the $n$-th corrections to be needed when $\alpha \gtrsim 10^{-5/n}$.

\section{Nondegenerate coupled fields}
\label{sec:NonDegCoupled}

Let us now consider the coupling between any two of the scalar, vector, axial gravitational and polar gravitational perturbations (excluding for the moment couplings between axial and polar gravitational perturbations, which will be the subject of Section~\ref{sec:DegCoupled} below).  We will show that $d^{ij}_{(k)}$ is zero for $i\neq j$ (i.e., that leading-order corrections induced by the couplings are quadratic) and that the number of coupled fields does not change the values of $d^{ii}_{(k)}=d_{(k)}$ or $e^{ijpq}_{(ks)}$.  We have produced an extensive list of the coefficients $e^{ijpq}_{(ks)}$ for $\ell = 2,\ldots,5$ and $k,s=0,\ldots,10$, which is available online~\cite{GRITJHU}.

The unperturbed QNM spectrum is the union of the spectra for each unperturbed field.  In this section we assume that these unperturbed spectra are nondegenerate. Corrections around the tensor QNM spectrum will be called tensor-led in the following. Similarly, corrections around the scalar (vector) QNM spectra are scalar- (vector-)led, respectively.  Spectra with multiple branches, corresponding to different fields, have been observed in extreme-mass ratio simulations and nonlinear BH mergers in Einstein-Maxwell theory~\cite{Johnston:1973cd,Johnston:1974vf,Zilhao:2012gp,Cardoso:2016olt}, Chern-Simons theory~\cite{Molina:2010fb}, and scalar Gauss-Bonnet gravity~\cite{Blazquez-Salcedo:2016enn,Witek:2018dmd}.  

The argument used in the derivation of Eqs.~\eqref{eq:omega1Uncoupled} and \eqref{eq:ksOmega2Uncoupled} can be generalized with the replacement $\alpha_{(k)}\to \alpha^{ij}_{(k)}$. The result is
\begin{align}
    \label{eq:omega1}
  d^{ij}_{(k)} =& -\frac{1}{\left.\partial_\omega {\cal S}\right|_{(\omega_0,0)}}
                  \left.\frac{\partial {\cal S}}{\partial\alpha_{ij}^{(k)}}\right|_{(\omega_0,0)}
\end{align}
and
\begin{align}
    \label{eq:omega2}
  e^{ijpq}_{(ks)} =& -\frac{1}{\left.\partial_\omega {\cal S}\right|_{(\omega_0,0)}}
                     \bigg(\frac{\partial^2}{\partial\alpha_{ij}^{(k)}\partial\alpha_{pq}^{(s)}}+d^{ij}_{(k)}\frac{\partial^2}{\partial\alpha_{pq}^{(s)}\partial\omega} \nonumber \\
    &\left. +d^{pq}_{(s)}\frac{\partial^2}{\partial\alpha_{ij}^{(k)}\partial\omega}+d^{ij}_{(k)}d^{pq}_{(s)}\frac{\partial^2}{\partial\omega^2}
      \bigg){\cal S}\right|_{(\omega_0,0)}
      \,.
\end{align}
While these equations look more complex than Eqs.~\eqref{eq:omega1Uncoupled} and \eqref{eq:ksOmega2Uncoupled}, they are entirely equivalent but expressed in full generality. 
Moreover, they reduce to the single-field case when $i=j=p=q$.

We now discuss some important subtleties in problems involving coupled fields. 

\subsection{Effect of $N$ coupled fields}

One may worry that if we consider the case of three fields and only couple two of them, the resulting $d^{ij}_{(k)}$ may differ from the case of only two fields.
This is because the roots of Eq.~\eqref{eq:detS} will remain the same when the number of fields increases, but the functional dependence of the determinant around the roots will not. 
Therefore, the derivatives in Eqs.~\eqref{eq:omega1} and~\eqref{eq:omega2} will be different depending on the number of fields we consider. 

Fortunately, the structure of Eq.~\eqref{eq:detS} together with the $\alpha_{(k)}^{ij}$ expansion implies that we only need to consider the corrections due to the roots of the $1\times1$ and $2\times2$ minors of $\mathbf{S}$ which contain the field corresponding to the spectra we perturb about.  Hence for each $\omega_0$, one $1\times1$ minor and $(N-1)$ $2\times2$ minors contribute corrections at quadratic order. This is shown in Appendix \ref{app:threeFields}. As a result, the value of $N$ plays no role in the calculation of $d^{ij}_{(k)}$ or $e^{ijpq}_{(ks)}$, beyond setting the range of the sums over the indices $i$ and $j$.

The $1\times1$ minor of the matrix $\mathbf{S}$ corresponds to the case of an uncoupled field.
Moreover, the $1\times1$ minor captures all effects of the additional potentials to quadratic order.
Hence, we can make the identification $d^{ii}_{(k)}=d_{(k)}$ and $e^{iiii}_{(ks)}=e_{(ks)}$, where $d_{(k)}$ and $e_{(ks)}$ are found from the uncoupled case in Section~\ref{sec:QuadraticUncoupled}. 

The roots of the $2\times2$ minors will give corrections due to field couplings up to quadratic order. 
We will now only consider the case of 2 coupled fields and examine the roots of these $2\times2$ minors.

\subsection{Coupling between nondegenerate spectra generates quadratic corrections}

Diagonal perturbation terms $\delta V_{ii}(r)$ generate linear and quadratic corrections to the QNM frequencies. On the contrary, perturbative couplings $\delta V_{ij}(r)$ only generate quadratic corrections, as long as the spectra of the unperturbed fields are nondegenerate. This is shown in Appendix \ref{app:threeFields} by examining the structure of $\mathbf{S}$.

While corrections to the QNM frequencies due to the coupling between two fields with nondegenerate spectra are quadratic in $\alpha$, the magnitude of these coupling-induced corrections when the values of $\alpha$ are specified need not be smaller than a given linear correction: for example, in dCS gravity $\alpha^{(k)}_{11} = \mathcal{O}(\epsilon)$ and $\alpha^{(k)}_{12} = \mathcal{O}(\epsilon^{1/2})$ (see e.g.~\cite{Kimura:2018nxk} and Sec.~\ref{sec:Examples}), so the two corrections are formally of the same order in $\epsilon$.

Since there are no linear corrections, Eq.~\eqref{eq:omega2} simplifies to
\begin{equation}
  e^{ijpq}_{(ks)} = -\left.\frac{1}{\partial_\omega {\cal S}}\right|_{(\omega_0,0)} \left.\bigg(\frac{\partial^2}{\partial\alpha_{ij}^{(k)}\partial\alpha_{pq}^{(s)}}\bigg){\cal S}\right|_{(\omega_0,0)}  \,.
\label{eq:omega2Simple}
\end{equation}
Some representative coefficients for scalar-axial gravitational and scalar-polar gravitational couplings are listed in Table~\ref{tab:quadraticCoupled}.

\begingroup
\setlength{\tabcolsep}{8pt} %
\begin{table}[]
\begin{tabular}{ccc}
\hline
\hline
    & Axial-Scalar  & Polar-Scalar \\
$k$ & $r_H\,e^{1221}_{(kk)}$ & $r_H\,e^{1221}_{(kk)}$ \\ 
\hline
\hline
 2 &  -0.0388 - 0.00196 i & -0.0386 -0.00135 i\\
 3 &  -0.0146 + 0.000930 i & -0.0155 + 0.00162 i\\
 4 &  -0.00567 - 0.000484 i & -0.00644 + 0.00000923 i\\
 5 &  -0.00228 - 0.00116 i & -0.00288 -0.000923 i\\
 10 &  0.000457 - 0.000387 i & 0.000318 - 0.000545 i \\
\hline
\hline
\end{tabular}
\caption{
Quadratic correction coefficients $e^{1221}_{(kk)}$ [cf. Eq.~\eqref{eq:omega2Simple}] for scalar-axial gravitational and scalar-polar gravitational couplings.
We show tensor-led $\ell=2$ corrections for a few selected values of $k$. The coefficients for $k,s=0,\ldots,10$ are available online~\cite{GRITJHU}.
}
\label{tab:quadraticCoupled}
\end{table}

We stress again that these results only hold when the two unperturbed fields are not isospectral. For example, in the derivation of Eqs.~\eqref{eq:omega1} and \eqref{eq:omega2} we have made the assumption that $\partial_\omega {\cal S}|_{(\omega_0,0)} \neq 0$, which does not hold when the union of the unperturbed spectra is degenerate at $\omega_0$.  This assumption rules out the important case of a coupling between the axial and polar gravitational perturbations, which are known to be isospectral. We now turn to the effect of couplings between fields with degenerate spectra.

\section{Degenerate coupled fields}
\label{sec:DegCoupled}

So far we made the assumption that the spectra of the coupled system in Eq.~\eqref{eq:waveEQ} are nondegenerate in the unperturbed case, i.e. when $\alpha_{ij}^{(k)}=0$ for all $i,\,j,\,k$.
This was used to obtain Eqs.~\eqref{eq:omega1} and \eqref{eq:omega2}, and also in Appendix \ref{app:threeFields}.
Unfortunately this assumption is not valid whenever there is a coupling between the axial and polar gravitational perturbations, because  
in GR the corresponding QNM frequencies are well known to be isospectral~\cite{Chandrasekhar:1975zza,Chandrasekhar:1985kt}.

In this section we show that couplings between degenerate spectra yield linear corrections to the QNM frequencies. 
Furthermore, the expansion \eqref{eq:OmegaExpansion} does not apply in this case. If all elements of $\delta \mathbf{V}$ are nonzero, the total first-order correction to the spectra is not the sum of the corrections found from each individual element of $\delta \mathbf{V}$, but it is rather a nonlinear combination of these corrections.

To illustrate the nature of the problem, let us start again with a simple example.  Consider the system defined by
\begin{align}
    \left(\frac{d^2}{dr_*^2} +\omega^2 - f V_0 \right) \phi_1 + \alpha Z \phi_2 = 0\,, \\
    \left(\frac{d^2}{dr_*^2} +\omega^2 - f V_0 \right) \phi_2 + \alpha Z \phi_1 = 0\,,
    \label{eq:SimpleCoupled}
\end{align}
for some potential $V_0$ and coupling $Z$. 
The QNM spectra of $\phi_1$ and $\phi_2$ are trivially degenerate for $\alpha=0$.
To see that corrections enter at linear order, diagonalize the system with the transformation 
\begin{align}
    \phi_1 = (\phi_+ + \phi_-)/2\,,\\
    \phi_2 = (\phi_+ - \phi_-)/2\,.
\end{align}
The resulting equations are then 
\begin{align}
    \left(\frac{d^2}{dr_*^2} +\omega^2 - f V_0 +  \alpha Z\right) \phi_+ = 0 \,, \\
    \left(\frac{d^2}{dr_*^2} +\omega^2 - f V_0 -  \alpha Z\right )\phi_- = 0\,.
\end{align}
It is clear that the corrections to the spectra will enter at linear order in $\alpha$ despite the initial perturbations being coupled, whereas the results of the previous section would imply that the correction should be quadratic in $\alpha$.

\begingroup
\setlength{\tabcolsep}{8pt} %
\begin{table}[]
\begin{tabular}{ccc}
\hline
\hline
$k$ & $\delta V_{-+}^{(k)}\delta V_{+-}^{(k)}$ \\ 
\hline
\hline
 2 &  0.0324 - 0.000828 i\\
 3 &  0.0313 -0.00321 i\\
 4 &  -0.000221 - 0.000215 i\\
 5 &  0.0000165 + 0.00000100 i\\
 10 & 0.00242216 + 0.00309852 i \\
\hline
\hline
\end{tabular}
\caption{
The product $\delta V_{-+}^{(k)}\delta V_{+-}^{(k)}$ defined in  Eqs.~(\ref{eq:deltaVkppDef}) and (\ref{eq:deltaVkpmDef}) for the case of axial-polar couplings for $\ell=2$.
The coefficients for $l=2$  and $k,s=0,\ldots,10$ are available online~\cite{GRITJHU}.
}
\label{tab:ExpValues}
\end{table}

In Appendix~\ref{app:Degenerateomega1derivation} we consider a general potential matrix of the form \eqref{eq:PotentialMatrix}. Assuming a QNM frequency expansion of the form 
\begin{equation}
    \omega = \omega_0 + \epsilon \omega_1,
\end{equation}
we show that
\begin{align}
    \omega_1 = \frac{ \delta V_{++} + \delta V_{--} \pm \sqrt{( \delta V_{++} - \delta  V_{--})^2 + 4 \delta V_{+-} \delta V_{-+}}}{2}\,.
\label{eq:DegLinearE1Main}
\end{align}
The coefficients  $\delta V_{+-}$ (for example) have the expansion
\begin{equation}
\label{eq:ExpectationExpansion}
    \delta V_{+-} = \sum_{k =0}^\infty a_{+-}^{(k)} \delta V^{(k)}_{+-} \,.
\end{equation}
Each of the factors $\delta V^{(k)}_{\pm\pm}$ and $\delta V^{(k)}_{\pm\mp}$ is related to the expectation value of the $k$th term in a power-series expansion of the potential perturbations:
see Eqs.~(\ref{eq:deltaVkppDef}) and (\ref{eq:deltaVkpmDef}) for their definitions.

\begin{figure*}[thbp] \includegraphics[width=0.75\columnwidth]{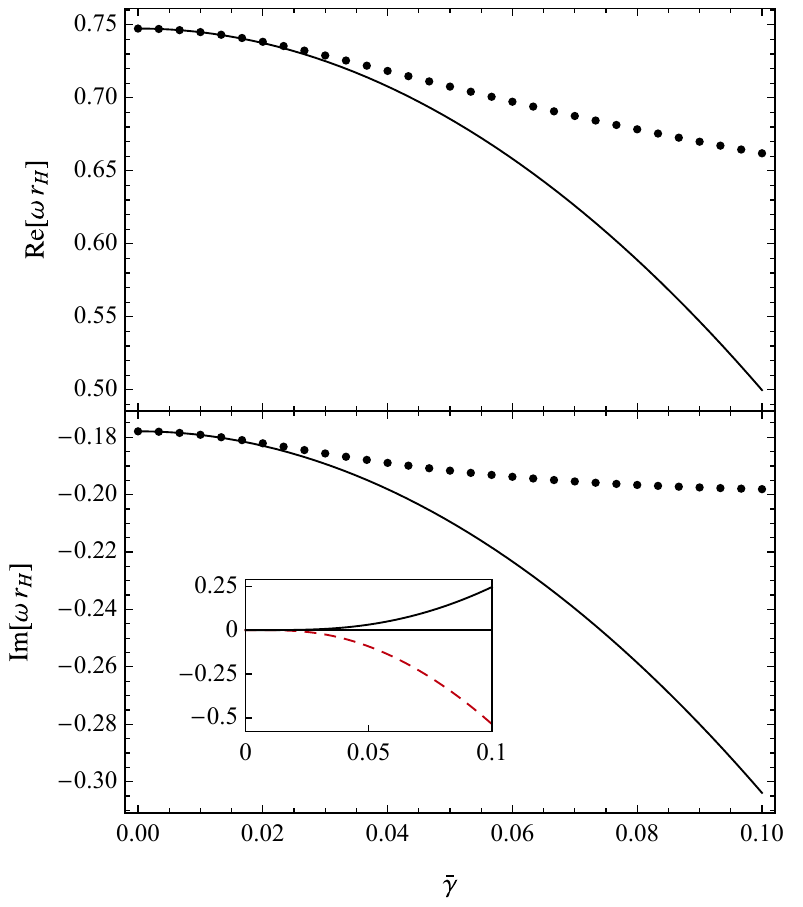}
  \includegraphics[width=0.75\columnwidth]{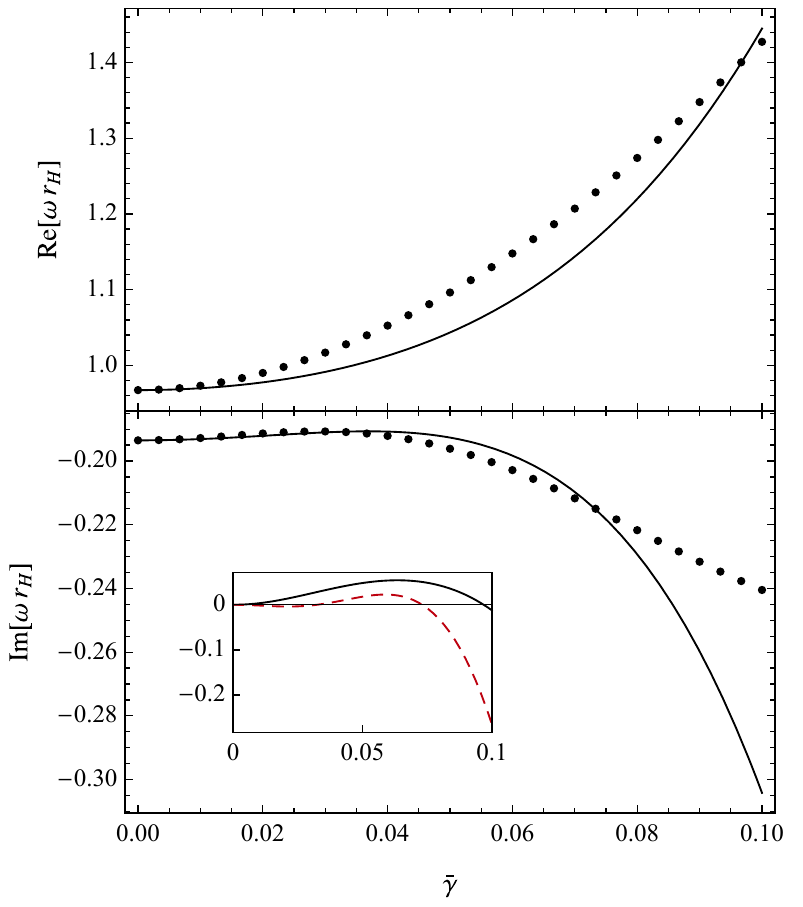}\\
  \caption{Fundamental axial $\ell=2$ QNM frequencies of Schwarzschild BHs in dCS gravity as function of $\bar\gamma$ for the tensor-led (left) and scalar-led (right) modes.  Solid lines refer to Eqs.~\eqref{dCSgrav} and \eqref{dCSscal}; bullets were computed through a direct integration method. The inset shows the relative difference between the two calculations for the real (solid black lines) and imaginary (dashed red lines) part of the modes.}
\label{fig:dCS}
\end{figure*}

One might have hoped that the corrections to the QNM spectra from the coupling of the two degenerate fields would allow for an expansion analogous to Eq.~\eqref{eq:OmegaExpansion}.  Indeed, in the absence of coupling, the argument of the square root in Eq.~(\ref{eq:DegLinearE1Main}) becomes a square, and we do recover a linear sum over single-field expectation values.  However, in general, the presence of couplings makes this relation nonlinear. Therefore we provide the values of the quantities $\delta V^{(k)}_{-+}\delta V^{(s)}_{+-}$ for $k,s=0,...10$ and $\ell=2\ldots5$ online~\cite{GRITJHU}. In Table~\ref{tab:ExpValues} we also list a small sample of values of $\delta V_{-+}^{(k)}\delta V_{+-}^{(k)}$ for $\ell=2$.  In order to find linear corrections to the QNM frequencies, these quantities must be plugged into Eqs.~\eqref{eq:DegLinearE1Main} and \eqref{eq:ExpectationExpansion}.  Note that $\delta V_{++}^{(k)}$ and $\delta V_{--}^{(k)}$ are just the coefficients $d_{(k)}$ for uncoupled (axial or polar) gravitational perturbations.

\section{Examples}
\label{sec:Examples}

For illustration, we now apply the formalism to compute QNM spectra for some classes of modified theories of gravity that are known to lead to coupled perturbation equations.
Specifically, we consider two models where the coupling is between scalar and tensor modes (dCS gravity~\cite{Kimura:2018nxk} and Horndeski gravity~\cite{Tattersall:2018nve}) and a model where the coupling is between axial and polar gravitational perturbations (the EFT inspired model~\cite{Cardoso:2018ptl} not considered in detail in Paper I), so that the background QNM spectra are degenerate. 

\subsection{Dynamical Chern-Simons gravity}

In dCS gravity, an effective low-energy theory with an additional scalar degree of freedom \cite{Alexander:2009tp},
nonspinning BHs are described by the Schwarzschild metric. The polar sector of gravitational perturbations is the same as in GR, whereas axial gravitational perturbations and scalar perturbations lead to a coupled system of the form \eqref{eq:waveEQ}~\cite{Cardoso:2009pk,Molina:2010fb} with the following potentials, in the notation of Eqs.~\eqref{eq:PotentialMatrix}, (\ref{eq:EffVm}) and (\ref{eq:EffVs}):
\begin{align}
    V_{11}&=V_-\,,\label{eq:potdcs1}\\
    V_{12}&=V_{21}= \frac{1}{ r_H^2}\frac{12}{\sqrt{\beta } r_H^2}\sqrt{\pi \frac{(\ell+2)!}{(\ell-2)!}}\left(\frac{r_H}{r}\right)^5\,,\\
    V_{22}&= V_{s=0} + \frac{1}{r_H^2}\frac{144 \pi  \ell(\ell+1)}{\beta r_H^4}\left(\frac{r_H}{r}\right)^8\,.\label{eq:potdcs2}
\end{align}
The parameter $\beta$ appearing in the dCS action has dimensions $[L]^{-4}$ and it sets the strength of the coupling, playing a role similar to the Brans-Dicke parameter $\omega_{\rm BD}$. It is useful to introduce a small dimensionless coupling parameter $\bar\gamma\equiv \beta^{-1/2}r_H^{-2}$ such that the equations decouple in the GR limit $\bar\gamma\to0$.

We first study how the parameter $\bar\gamma$ modifies the tensor-led mode. 
Using Eq.~\eqref{eq:OmegaExpansion} and reading off the relevant coefficients from the potentials~\eqref{eq:potdcs1}--\eqref{eq:potdcs2} we find
\begin{equation}
  \label{dCSgrav}
  \omega = \omega_0 + e_{(55)}^{1221}\left({12}\bar\gamma\sqrt{\pi \frac{(\ell+2)!}{(\ell-2)!}}\right)^2\,.
\end{equation}

Proceeding similarly for the scalar-led mode, we find
\begin{align}
  \label{dCSscal}
    \omega &= \omega_0 + 2 d_{(8)}{144 \pi  \ell(\ell+1)}\bar\gamma^2
    + e_{(88)}\left[{144 \pi  \ell(\ell+1)}\bar\gamma^2\right]^2  \nonumber \\
    &+ e_{(55)}^{1221}\left({12}\bar\gamma\sqrt{\pi \frac{(\ell+2)!}{(\ell-2)!}}\right)^2\,.
\end{align}
These expressions illustrate the importance of specifying the form of the coupling: since the off-diagonal terms $V_{12}$ and $V_{21}$ are proportional to $\beta^{-1/2}$, coupling-induced corrections end up being of the same order as the corrections due to $\delta V_{22}$.

Tensor-led and scalar-led dCS QNM frequencies have been previosuly computed using various methods~\cite{Molina:2010fb,Okounkova:2019dfo}.  In Fig.~\ref{fig:dCS} we compare Eqs.~\eqref{dCSgrav} and~\eqref{dCSscal} (solid lines) with a numerical QNM calculation based on the direct integration of the perturbed field equations (bullets) for the fundamental $\ell=2$ mode. The left panel refers to tensor-led modes, while the right panel refers to scalar-led modes. In the inset we show the relative difference between the quadratic expansions of Eqs.~\eqref{dCSgrav} and~\eqref{dCSscal} and the numerical calculation for the real (solid) and imaginary (dashed) parts of the QNM frequencies.

\subsection{Horndeski gravity}

The perturbations of scalar and tensor fields on a Schwarzschild background in Horndeski gravity were studied 
in~\cite{Tattersall:2018nve} . The even-parity perturbation equations can be separated through field redefinitions, 
but there is coupling between even gravitational and scalar perturbations~\cite{Tattersall:2017erk}.
    
The master equation for the scalar-led modes takes the form 
\begin{equation}
    \frac{d^2\phi}{dr_*^2} + \left[ \omega^2 - f \left(V_{s=0} +  \mu^2+\frac{\ell(\ell+1)}{r^2}f \Gamma\right)\right]\phi=0\,.\label{hornscalar}
\end{equation}

The change to the spectrum is determined by an ``effective mass'' parameter $\mu$ and by a second parameter 
$\Gamma$, built out of the background values (denoted here by overbars) of the free functions appearing in the 
Horndeski action~\cite{Horndeski:1974wa,Kobayashi:2011nu}:
\begin{align}
    \mu^2 &= \frac{-\overline{G}_{2\phi\phi}}{3\overline{G}^2_{4\phi}+\overline{G}_{2X}-2\overline{G}_{3\phi}}\,, \\
    \Gamma &= \frac{8\overline{G}_{4X}}{3\overline{G}^2_{4\phi}+\overline{G}_{2X}-2\overline{G}_{3\phi}}\,.
\end{align}

Let $\omega_0$ belong to the spectrum of $V_{s=0}$. Then the scalar-led frequencies can be approximated as
\begin{align}
    \omega &\approx \omega_0 + d_{(0)}\mu^2 + [d_{(2)} + r_H d_{(3)}]\ell(\ell+1)\Gamma  + \frac{1}{2}e_{(00)}\mu^4 \nonumber \\
    &+ \frac{1}{2}[e_{(22)} + 2 r_H e_{(23)} + r_H^2 e_{(33)}][\ell(\ell+1)\Gamma]^2 \nonumber \\
    &+[e_{(02)}+r_H e_{(03)}]\ell(\ell+1)\mu^2\Gamma\,.
\end{align}
As noted in \cite{Tattersall:2018nve}, for the class of Horndeski theories in which the GW speed propagation satisfies $c_{\rm T}=1$, the 
$\Gamma$ factor vanishes identically. In this case Eq.~\eqref{hornscalar} depends on a single free parameter, given by the effective mass 
$\mu$, and the scalar-led frequencies simply reduce to 
\begin{align}
    \omega &\approx \omega_0 + d_{(0)}\mu^2 + \frac{1}{2}e_{(00)}\mu^4  \ .\label{hordct1} 
\end{align}
We compute the values of $\omega$ for the $\ell=2$ scalar mode in Horndeski gravity using a direct integration method as a function of the effective mass $\mu$. In Fig.~\ref{fig:horndeski} we plot numerical results (bullets) against the results obtained from Eq.~\eqref{hordct1} (solid lines). The two approaches are in excellent agreement, with relative deviations $\Delta\omega=(\omega^{\rm dir}-\omega^{\rm fit})/\omega^{\rm fit} \lesssim 10^{-4}$ for both the real and imaginary parts within the range of masses we consider.

\begin{figure}%
  \includegraphics[width=\columnwidth]{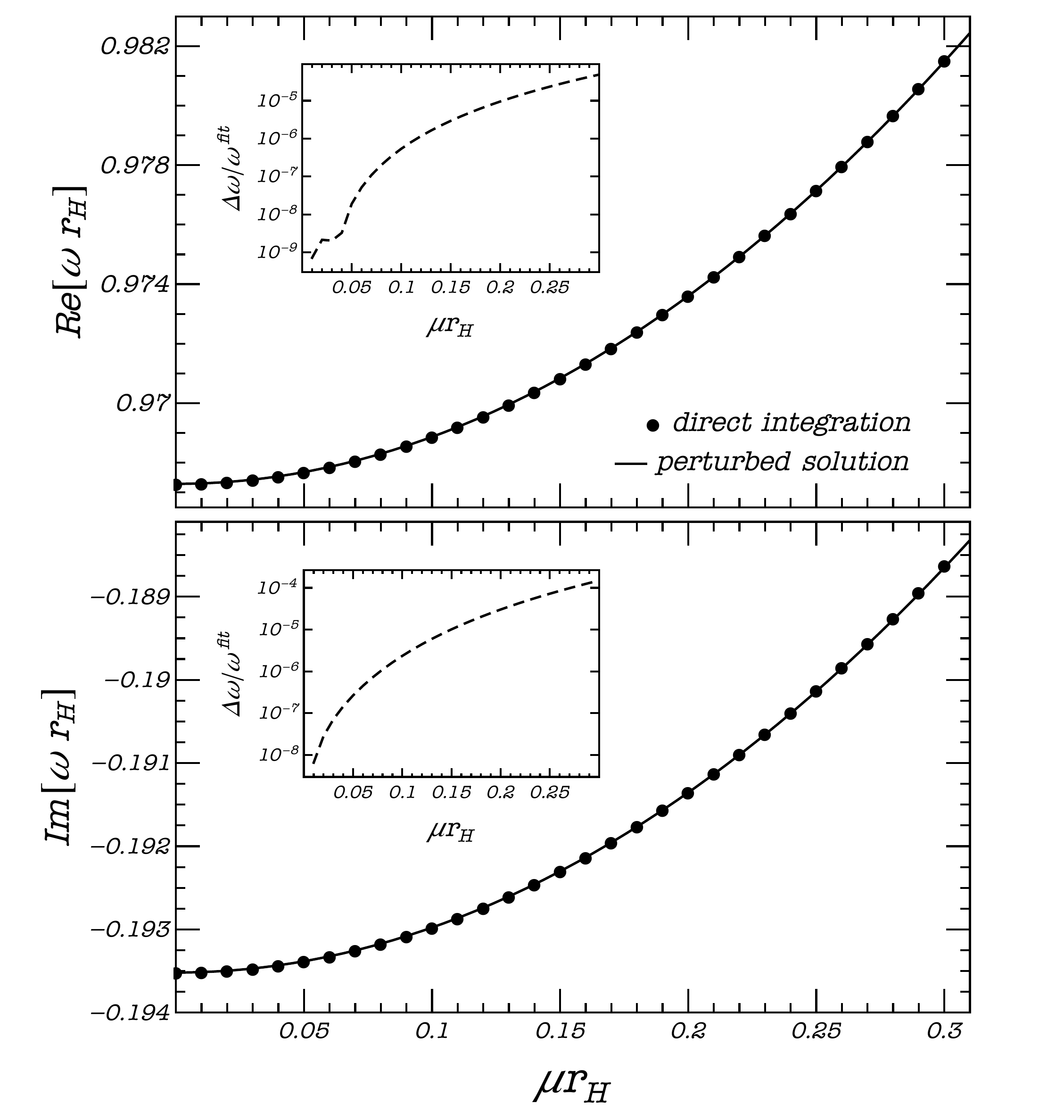}%
  \caption{Real and imaginary part of the $\ell=2$ mode of the scalar-led mode in Horndeski gravity.  The numerical results obtained 
  through direct integration (bullet points) are in excellent agreement with Eq.~\eqref{hordct1} (solid line). The insets in each panel 
  show the relative deviation between the two calculations.}%
\label{fig:horndeski}%
\end{figure}

\subsection{An effective field theory model}

One of the EFT models considered in~\cite{Cardoso:2018ptl} couples the axial and polar gravitational perturbations.  The wave equations map onto Eq.~\eqref{eq:waveEQ} with
\begin{align}
    V_{11} &= V_+\,,\\
    V_{22} &= V_-\,,\\
    V_{12} &= V_{21} = \epsilon V(r)\,,
\end{align}
where $V_-$ and $V_+$ were defined in Eqs.~(\ref{eq:EffVm}) and (\ref{eq:EffVp}), $V(r)$ can be found in Appendix A of~\cite{Cardoso:2018ptl}, and the small dimensionless parameter $\epsilon$ is inversely related to the UV cutoff scale of the EFT.

The spectra of $V_+$ and $V_-$ are degenerate when $\epsilon =0$, so the coupling should induce linear corrections, as discussed in Section~\ref{sec:DegCoupled}. This expectation is verified in Figure~\ref{fig:e3EFT}. There we consider perturbations about the fundamental QNM frequency with $\ell=2$, and we show how the two branches of QNM frequencies -- corresponding to the two possible sign choices in Eq.~(\ref{eq:DegLinearE1Main}) -- change with $\epsilon$.
Note that the coupled QNM frequencies computed by direct integration in Fig.~\ref{fig:horndeski} are novel results that were not presented in~\cite{Cardoso:2018ptl}.

Fitting the two branches of QNM frequencies to a seventh-order polynomial in $\epsilon$ gives a linear coefficient $(-0.1479-0.2729 i)r_H$ 
for the solid black branch, and $(0.1480+0.2719 i)r_H$ for the dashed red branch.
The linear corrections are nonzero, as they should, because the uncoupled axial and polar spectra are degenerate.

\begin{figure}[t]
  \includegraphics[width=\columnwidth]{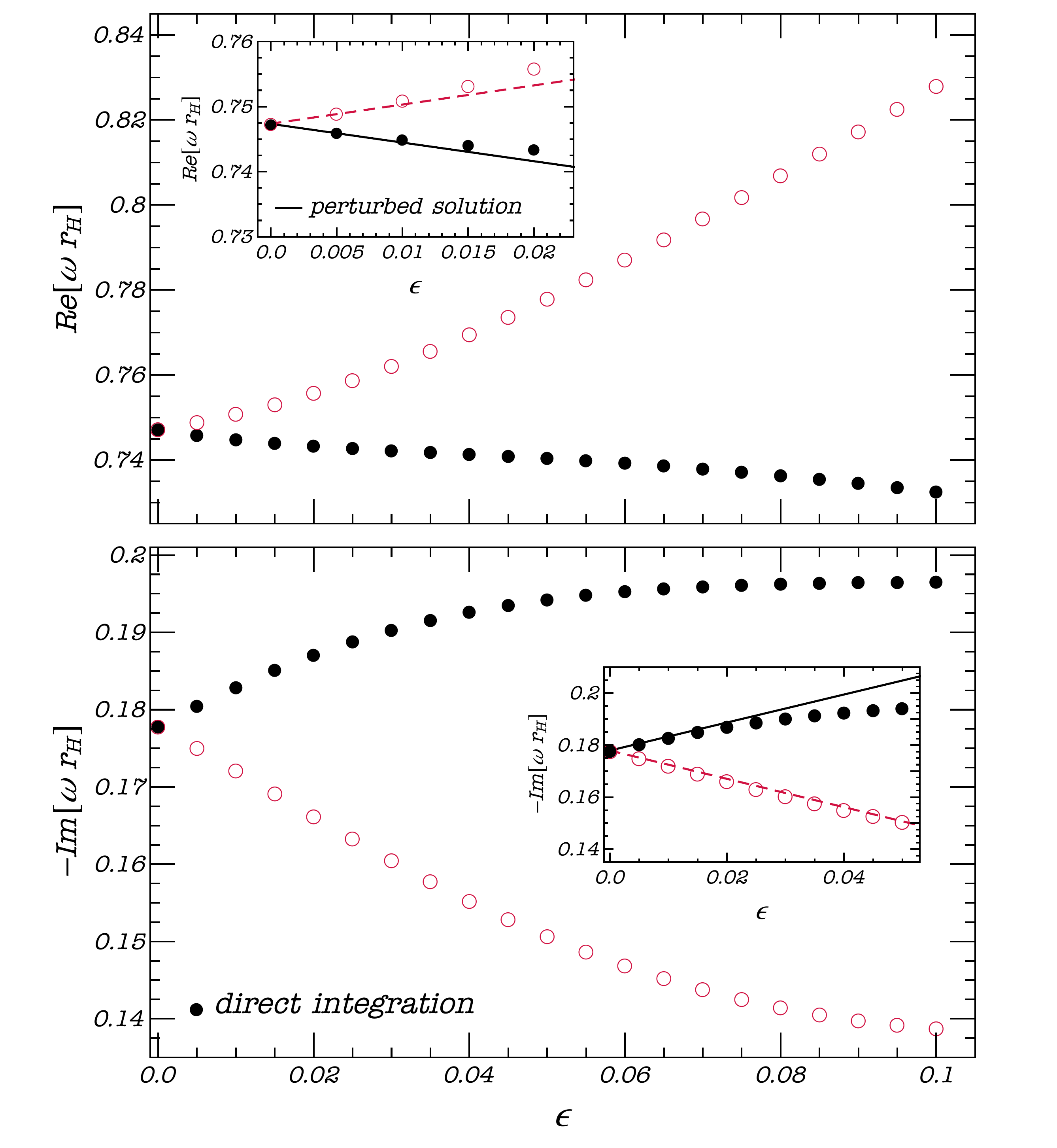}\\
  \caption{
  Perturbations of the fundamental QNM frequencies with $\ell=2$ for axial and polar gravitational perturbations, coupled according to the EFT model of~\cite{Cardoso:2018ptl}. 
  The bullet points correspond to frequency values obtained through a direct integration method. 
  The insets show a zoom for small $\epsilon$ where the behavior of $\omega$ deviate from the linear trend, and therefore from the linear approximation given by Eq.~\eqref{eq:DegLinearE1Main}.}
\label{fig:e3EFT}
\end{figure}

\section{Conclusions and a computational recipe}
\label{sec:Conclusion}

We have extended the formalism of Paper I to compute QNM frequencies of coupled fields as long as the perturbations they induce are small deviations from the perturbation equations for the Schwarzschild geometry in GR. Our main result is a convenient, ready-to-use recipe to compute QNM frequencies at quadratic order in the perturbations. We crucially allow for the possibility of coupling between the master equations. First-order  (friction-like) terms in the field derivatives can be accommodated through field redefinitions (cf. Appendix~\ref{app:NoFriction}).

We have found the expansion of the QNM frequencies for uncoupled wave equations to quadratic order. Perhaps our most interesting findings concern the QNM spectra of coupled fields.
When the coupling occurs between fields with nondegenerate spectra at zero order in the perturbations, linear-order corrections to the QNM frequencies vanish.  However, when the coupling occurs between fields with degenerate spectra, the QNM frequency corrections are {\em linear} in the perturbations.

Our results significantly simplify the task of computing QNM frequencies in any modified theory of gravity, or any theory allowing for additional fields. The general recipe for this calculation can be summarized as follows:

\begin{itemize}
\item[1)] Derive the master equations for the perturbation variables in the given theory;
  
\item[2)] Eliminate first-order (friction-like) terms in the field derivatives through field redefinitions, as described in Appendix~\ref{app:NoFriction};

\item[3)] Identify the relevant coefficients $\alpha_{ij}^{(k)}$ in the perturbed potentials $\delta V_{ij}$ [Eq.~(\ref{eq:PotentialMatrix})] appearing in the general coupled system of Eq.~(\ref{eq:waveEQ}); if these coefficients are frequency-dependent, compute their frequency derivative $\alpha'{}_{pq}^{(s)}$.

\item[4a)] If any two unperturbed spectra are nondegenerate, compute corrections to the QNM frequencies by simple multiplications and additions using Eq.~(\ref{eq:OmegaExpansion}) and the tabulated values of $d^{ij}_{(k)}$ and $e^{ijpq}_{(ks)}$, which are available online~\cite{GRITJHU}.

\item[4b)] If any two unperturbed spectra are degenerate, compute corrections to the QNM frequencies using Eq.~(\ref{eq:DegLinearE1Main}) and the tabulated values of $\delta V^{(k)}_{\pm\pm}$ and $\delta V^{(k)}_{\pm\mp}$  defined in Eqs.~(\ref{eq:deltaVkppDef}) and (\ref{eq:deltaVkpmDef}), which are also available online~\cite{GRITJHU}.
\end{itemize}

In Sec.~\ref{sec:Examples} we illustrate this procedure for three classes of modified theories of gravity leading to coupled perturbation equations: two models coupling the scalar and tensor modes (dCS gravity~\cite{Kimura:2018nxk} and Horndeski gravity~\cite{Tattersall:2018nve}) and an EFT model coupling the axial and polar gravitational perturbations~\cite{Cardoso:2018ptl}, where the background QNM spectra are degenerate. 

While our expansion is theory-agnostic, we make assumptions about the effect of the modified gravity theory: the background should be perturbatively close to the Schwarzschild metric, and the corrections to the ``ordinary'' potentials in the GR master equations should be amenable to a power-series expansion in inverse powers of the radial variable.

This results by construction in small corrections to the GR QNM spectra~\eqref{eq:OmegaExpansion}.  In general, as discussed in Paper I, new nonperturbative frequencies (e.g., quasibound states emerging from zero frequency for massive scalars) may appear in the spectrum, and these are not captured by our formalism.

The assumption that the background is only perturbatively different from the Schwarzschild solution is slightly less restrictive than one might think. For example, in Paper I we showed that slowly rotating Kerr BHs can be accommodated within the formalism. Recent work on higher-derivative corrections to the Kerr geometry~\cite{Cano:2019ore} and on QNM frequencies of rotating solutions for small coupling~\cite{Zimmerman:2014aha,Mark:2014aja} may allow us to make progress on the calculation of QNMs in modified gravity for rotating BH remnants, such as those observed by the LIGO/Virgo collaboration~\cite{LIGOScientific:2018mvr}. Related attempts at parametrizing deviations from the Kerr QNM spectrum~\cite{Glampedakis:2017dvb,Glampedakis:2017cgd}  made use of the connection between the stability of null geodesics and QNMs~\cite{1971ApJ...170L.105P,1972ApJ...172L..95G,Cardoso:2008bp}. Our formalism may help to clarify the conditions under which this ``geodesic correspondence'' applies~\cite{Konoplya:2017wot,Glampedakis:2019dqh}.


%
\noindent{\bf{\em Acknowledgments.}}
We thank Macarena Lagos, Oliver Tattersall and Aaron Zimmerman for useful discussions.
E.B. and R.M. are supported by NSF Grant No. PHY-1841464, NSF Grant No. AST-1841358, NSF-XSEDE Grant No. PHY-090003, and NASA ATP Grant No. 17-ATP17-0225.
V.C. acknowledges financial support provided under the European Union's H2020 ERC 
Consolidator Grant ``Matter and strong-field gravity: New frontiers in Einstein's 
theory'' grant agreement no. MaGRaTh--646597.
This project has received funding from the European Union's Horizon 2020 research and innovation programme under the Marie Sklodowska-Curie grant agreement No 690904.
A.M. acknowledge support from the Amaldi Research Center funded by the MIUR program ``Dipartimento di Eccellenza'' (CUP: B81I18001170001). 
We acknowledge financial support provided by FCT/Portugal through grant PTDC/MAT-APL/30043/2017.
We acknowledge the SDSC Comet and TACC Stampede2 clusters through NSF-XSEDE Award Nos. PHY-090003,
as well as MareNostrum and the technical support provided by Barcelona Supercomputing Center (AECT-2018-1-0003).
The authors would like to acknowledge networking support by the GWverse COST Action 
CA16104, ``Black holes, gravitational waves and fundamental physics.''
We acknowledge support from the Amaldi Research Center funded by the 
MIUR program ``Dipartimento di Eccellenza''~(CUP: B81I18001170001).
C.F.B.M. thanks the Conselho Nacional de Desenvolvimento Científico e
Tecnológico (CNPq) for financial support,
 the Johns Hopkins University for kind hospitality during the preparation
of this work and the American Physical Society which funded
the visit through the International Research Travel Award Program.
The authors thankfully acknowledge the computer resources, technical expertise and assistance provided by CENTRA/IST. Computations were performed at the cluster “Baltasar-Sete-S\'ois” and supported by the H2020 ERC Consolidator Grant ``Matter and strong field gravity: New frontiers in Einstein's theory'' grant agreement no. MaGRaTh-646597.

\appendix
\section{Friction-like terms}
\label{app:NoFriction}

The master wave equations for coupled fields can, in general, contain
``friction-like'' terms (i.e., terms which are linear in derivatives of the perturbation variable). Ref.~\cite{Tattersall:2017erk} lists several examples\footnote{See e.g. their Eq.~(60) for even-parity scalar-tensor perturbations; Eqs.~(88) and (89) for odd-parity vector-tensor perturbations; Eqs.~(94)--(97) for even-parity vector-tensor perturbations; and Eq.~(104) for even-parity massive Proca perturbations~\cite{Tattersall:2017erk}.} of coupled wave equations of this type.

In this appendix we show that a matrix $\mathbf{Z}$ of friction-like terms (proportional to the first radial derivative of the wave function) can always be reabsorbed in the potential matrix $\mathbf{V}$ through suitable field redefinitions.

\subsection{Single-field case}

For simplicity, let us first consider adding a friction-like term to the master equation:
\begin{align}
    \frac{d^2 }{dr_*^2} + \epsilon Z(r_*)\frac{d\phi}{dr_*}  + (\omega^2 - V) \phi &= 0\,,
\end{align}
for some function $Z(r_*)$ and small parameter $\epsilon$ (note that in this appendix, and here only, we redefine $fV\to V$ to simplify the notation).
A field redefinition
\begin{align}
     \phi = \xi e^{-\frac{\epsilon}{2} \int^{r_*} Z(t) dt} \,,
\end{align}
is sufficient to remove the term proportional to the first derivative~\cite{ArfkenWeber}.
The resulting field equation is
\begin{equation}
    \frac{d^2\xi}{dr_*^2} + \left(\omega^2 - V - \frac{1}{2}\epsilon Z' - \frac{1}{4}\epsilon^2 Z^2\right) \xi =0\,,
    \label{eq:singleFieldFrictionRemoved}
  \end{equation}
where the last term in parentheses (proportional to $\epsilon^2 Z^2$) can be ignored in our perturbative framework.  
Therefore we can use the formalism of the main text by a suitable redefinition of the radial wave function and of the radial potential. We will now extend this idea to the case of multiple, coupled fields.

\subsection{Coupled case}

Let us now add a matrix of friction-like terms to a coupled set of master equations: 
\begin{align}
\label{eq:frictionMaster}
    \frac{d^2 {\bm\Phi} }{dr_*^2} + \epsilon  \bm Z (r_*) \frac{d {\bm\Phi} }{dr_*}  +  \bm V (r_*) {\bm\Phi} = 0\,,
\end{align}
for some matrix of functions $\mathbf{Z}(r_*)$ and small parameter $\epsilon$. 
For simplicity, here and below we redefine $\omega^2-V_{1,2}\to V_{1,2}$, we use primes for derivatives with respect to $r_*$, and we consider the case of only two fields.

We make a field redefinition
\begin{equation}
     {\bm\Phi} = \left(\mathbf{1}-\frac{\epsilon}{2}\int^{r_*} \bm Z (t) dt \right) {\bm \chi} \,,
\end{equation}
and we multiply the resulting equations on the left by 
\begin{equation}
  \left(\mathbf{1}+\frac{\epsilon}{2}\int^{r_*}  \bm Z (t) dt \right).
\end{equation}
This yields
\begin{align}
\label{eq:frictionFinal}
    \frac{d^2 {\bm\chi} }{dr_*^2} + (\bm V + \epsilon \bm W) {\bm\chi} = 0\,,
\end{align}
where $\bm W$ is a matrix with elements
\begin{align}
    W_{11} &= \frac{1}{2}\left( V_{21}\int^{r_*}Z_{12}dr'_* - V_{12}\int^{r_*}Z_{21}dr'_* - Z_{11}'\right)\,,
             \label{eq:W11}
    \\
    W_{22} &= \frac{1}{2}\left( V_{12}\int^{r_*}Z_{21}dr'_* - V_{21}\int^{r_*}Z_{12}dr'_* - Z_{22}'\right)\,,
             \label{eq:W22}
  \\
    W_{12} &= \frac{1}{2}\bigg( 
    V_{12}\int^{r_*}(Z_{11}-Z_{22})dr'_*  \nonumber \\
    &+(V_{22}-V_{11})\int^{r_*}Z_{12}dr'_* - Z_{12}'
    \bigg)\,,
    \\
    W_{21} &= \frac{1}{2}\bigg( 
    V_{21}\int^{r_*}(Z_{22}-Z_{11})dr'_*  \nonumber \\
    &+(V_{11}-V_{22})\int^{r_*}Z_{21}dr'_* - Z_{12}'
      \bigg)\,.
      \label{eq:W21}
\end{align}
As a result, the field equations no longer depend on the first derivative of any field. When $\bf V$ and $\bf Z$ are diagonal we recover Eq.~(\ref{eq:singleFieldFrictionRemoved}) at linear order in $\epsilon$, and when $\epsilon=0$ we recover the unperturbed QNM spectrum.
  
In general, the removal of friction-like terms introduces new potentials. 
For these new potentials to fit into our perturbative formalism, all new contributions must vanish at the horizon and (at most) tend to a constant at infinity. 
One parametrization that would satisfy these requirements is 
\begin{equation}
\epsilon Z_{ij} = \frac{1}{r_H}\left(1 -\frac{r_H}{r}\right)^2\sum_{s\geq2}\beta^{(s)}_{ij}\frac{r_H^s}{r^s}
\end{equation}
for small parameters $\beta^{(s)}_{ij}$, which can be mapped to the parameters $\alpha^{(k)}_{ij}$ discussed in the body of the paper. 
The summation must start at $s=2$, so that the integrals in Eqs.~(\ref{eq:W11})--(\ref{eq:W21}) do not diverge as $r_*\to \infty$.

\section{Expansion of $\omega$}
\label{app:omegaExpansion}

The coupling parameters $\alpha^{(k)}_{ij}$ appearing in Eq.~(\ref{eq:PotentialMatrix}) can, in general, depend on the frequency $\omega$. The computation of the QNM frequency itself depends on the couplings, so a nontrivial $\omega$-dependence of the couplings will affect the QNMs.
Consider the implicit expansion of $\omega$ about the GR value $\omega_0$:
\begin{align}
    \omega &= \omega_0 + \alpha^{(k)}_{ij}(\omega) d_{(k)}^{ij} %
             + \frac{1}{2}\alpha^{(k)}_{ij}(\omega)\alpha^{(s)}_{pq}(\omega) e_{(ks)}^{ijpq} + \mathcal{O}(\alpha^3)\,.
    \label{eq:omegaAlphaOmegaExpansion}
\end{align}
One may also Taylor expand $\alpha^{k}_{ij}(\omega)$ about $\omega_0$, 
\begin{align}
    \alpha^{(k)}_{ij}(\omega) &= \alpha^{(k)}_{ij}|_{\omega_0} + \alpha'{}^{(k)}_{ij}|_{\omega_0}(\omega-\omega_0) + \mathcal{O}(\alpha^2)\,.
    \label{eq:alphaOmegaAlphaExpansion}
\end{align}
We assume that $\alpha^{(k)}_{ij}$ and all of its derivatives evaluated at $\omega_0$ are small, so that we can consider them to be of the same order. 
By substituting \eqref{eq:alphaOmegaAlphaExpansion} into \eqref{eq:omegaAlphaOmegaExpansion} we find 
\begin{align}
    \omega &\approx \omega_0 + [\alpha^{(k)}_{ij}|_{\omega_0} + \alpha'{}^{(k)}_{ij}|_{\omega_0}(\omega-\omega_0)] d_{(k)}^{ij} \nonumber \\
    &+ \frac{1}{2}\alpha^{(k)}_{ij}|_{\omega_0}\alpha^{(s)}_{pq}|_{\omega_0} e_{(ks)}^{ijpq} + \mathcal{O}(\alpha^3)\,.
    \label{eq:omegaAlphaOmegaExpansion2}
\end{align}
Finally, substitute \eqref{eq:omegaAlphaOmegaExpansion2} into itself to get the result quoted in Eq.~(\ref{eq:OmegaExpansion}):
\begin{align}
\omega &\approx \omega_0 + \alpha^{(k)}_{ij}|_{\omega_0}d_{(k)}^{ij} + \alpha'^{(k)}_{ij}|_{\omega_0}\alpha^{(s)}_{pq}|_{\omega_0}d_{(k)}^{ij} d_{(s)}^{pq} \nonumber \\
    &+ \frac{1}{2}\alpha^{(k)}_{ij}|_{\omega_0}\alpha^{(s)}_{pq}|_{\omega_0} e_{(ks)}^{ijpq} + \mathcal{O}(\alpha^3)\,.
\end{align}

\section{The case of three fields}
\label{app:threeFields}

Let us schematically write the coupled master equations for three fields as
\begin{align}
  \left( {\begin{array}{ccc}
   L_1 + \delta V_{11} & \delta V_{12} & \delta V_{13}\\
   \delta V_{21} & L_2 + \delta V_{22} & \delta V_{23}\\
   \delta V_{31} & \delta V_{32} & L_3 + \delta V_{33}\\
  \end{array} } \right)
  \left( {\begin{array}{c} \phi \\ \psi \\ \chi \\ \end{array} } \right) %
  =
  \omega^2 \left( {\begin{array}{c} \phi\\ \psi \\ \chi \\ \end{array} } \right)
\end{align}
for some linear operators $L_{1,2,3}$, perturbative potentials $\delta V_{ij}=\alpha_{ij} \bar V_{ij}$ and small parameters $\alpha_{ij}$ (in this appendix we slightly change the notation to minimize clutter). 

We assume that the spectrum of $L_1$ is nondegenerate with the spectra of both $L_2$ and $L_3$.
Expanding the fields in powers of $\alpha_{ij}$ to linear order we find
\begin{align}
    \phi &= \phi_0 + \alpha_{11} \phi_{11} + \alpha_{12} \phi_{12} + \alpha_{13} \phi_{13}\,,\\
    \psi &= \psi_0 + \alpha_{21} \psi_{21} + \alpha_{22} \psi_{22} + \alpha_{23} \psi_{23}\,,\\
    \chi &= \chi_0 + \alpha_{31} \chi_{31} + \alpha_{32} \chi_{32} + \alpha_{33} \chi_{33}\,.
\end{align}
Note that all of the first-order fields are only functions of the uncoupled zeroth-order fields.

Let us write down explicitly some of the field equations:
\begin{align}
  \label{eq:phi0FieldEq}
  \omega^2 \phi_0 &= L_1 \phi_0 \,,\\
    \label{eq:phi12FieldEq}
    \omega^2 \phi_{12} &= L_1 \phi_{12} + \bar V_{12} \psi_0 \,,\\
     \label{eq:psi23FieldEq}
    \omega^2 \psi_{23} &= L_1 \psi_{23} + \bar V_{23} \chi_0 \,.
\end{align}

Recall from Section~\ref{sec:numerics} that we numerically integrate the system of equations order by order with respect to a diagonal basis of initial values at the horizon.
We also need to specify the value of each perturbation at the horizon, which we set to zero for all but the zeroth-order component of each field, so when $\alpha_{ij}=0$ the perturbations are not excited. 
We will denote by $N^{\psi}(\phi_{12})$, for example, the numerically integrated solution to  Eq.~\eqref{eq:phi12FieldEq}.
In this notation, the superscript denotes which field is excited at the horizon during the integration. 

The purpose of this appendix is to show that the perturbed spectrum of $L_1$ is calculated using Eqs.~\eqref{eq:omega1} and \eqref{eq:omega2}, where the determinant ${\cal S}$ refers to the $1\times1$ or $2\times2$ minors of the $3\times3$ matrix
\begin{equation}
\label{eq:3detS}
   \mathbf{S} = \left( {\begin{array}{ccc}
   N^\phi(\phi) & N^\phi(\psi) & N^\phi(\chi)\\
   N^\psi(\phi) & N^\psi(\psi) & N^\psi(\chi)\\
   N^\chi(\phi) & N^\chi(\psi) & N^\chi(\chi)\\
  \end{array} } \right)
\end{equation}
which contain $N^\phi(\phi)$. The proof goes as follows.

Let $\omega$ be in the spectrum of $\phi_0$, so that $N^\phi(\phi_0)=0$: cf. Eq.~(\ref{eq:phi0FieldEq}). Then we can conclude that:

\begin{itemize}
\item[(i)]
$N^{\phi}(\psi_0) = N^{\phi}(\chi_0) = 0$, $N^{\psi}(\phi_0) = N^{\psi}(\chi_0) =0$ and $N^{\chi}(\phi_0) = N^{\chi}(\psi_0) = 0$, as the leading order is not excited by our choice of basis.

\item[(ii)]
$N^{\phi}(\phi_{12}) = N^{\phi}(\phi_{13})=0$: this is because the equations for $\phi_{12}$ and $\phi_{13}$ are sourced by $\psi_{0}$ and $\chi_{0}$ [cf. Eq.~(\ref{eq:phi12FieldEq})], which are zero by our choice of basis.

\item[(iii)]
$N^{\phi}(\psi_{23}) = 0$, and similar relations apply under permutations $\phi\to\psi\to\chi$. 
This is because the equations for $\psi_{23}$ is sourced by $\chi_{0}$ [cf. Eq.~(\ref{eq:psi23FieldEq})], which is zero by our choice of basis.
\end{itemize}

In conclusion, the matrix $\mathbf{S}$ reduces to
\begin{align}
   \left( {\begin{array}{ccc}
   \alpha _{11} N^\phi(\phi_{11}) & \alpha_{21} N^\phi(\psi_{21}) & \alpha_{31} N^\phi(\chi_{31})\\
   \alpha_{12} N^\psi(\phi_{12}) & N^\psi(\psi_{0}) + \alpha _{22} N^\psi(\psi_{22}) &  \alpha_{32} N^\psi(\chi_{32})\\
   \alpha_{13} N^\chi(\phi_{13}) & \alpha_{23} N^\chi(\psi_{23}) & N^\chi(\chi_{0}) + \alpha _{33} N^\chi(\chi_{33})\\
  \end{array} } \right)\,,
\end{align}
and its determinant reads
\begin{align}
\label{eq:detS33Expansion}
    {\cal S} &= \alpha_{11} N^{\phi}(\phi_{11})N^{\psi}(\psi_{0})N^{\chi}(\chi_{0}) \nonumber \\
    &-\alpha_{12} \alpha_{21} N^\phi(\psi_{21}) N^\psi(\phi_{12}) N^\chi(\chi_{0}) \nonumber \\
    &-\alpha_{13} \alpha_{31} N^\chi(\phi_{13}) N^\phi(\chi_{31}) N^\psi(\psi_{0}) \nonumber \\
    &+\alpha_{11} \alpha_{33} N^\phi(\phi_{11}) N^\chi(\chi_{33}) N^\psi(\psi_{0}) \nonumber \\
    &-\alpha_{11} \alpha_{22} N^\phi(\phi_{11}) N^\psi(\psi_{22}) N^\chi(\chi_{0}) \nonumber \\
    &+\mathcal{O}(\alpha^3)=0\,.
\end{align}

The $\alpha_{ij}$ coefficients of each term can be varied independently, so each term in this sum must vanish.  Note also that $N^\psi(\psi_0)\neq 0$ and $N^\chi(\chi_0)\neq 0$, as we have assumed that $L_1$ is not degenerate with $L_2$ and $L_3$.  The first three lines in this sum give the corrections to linear and quadratic order, and are explicitly those formed from the $1\times1$ and the $2\times2$ minors of $\mathbf{S}$ containing the $(1,\,1)$ element, because we perturb around the spectrum of $L_1$.  The last two lines vanish identically, because -- recalling that $N^\psi(\psi_0)\neq 0$ and $N^\chi(\chi_0)\neq 0$ -- the first line implies $N^\phi(\phi_{11})=0$.

In conclusion: adding a third field does not affect our expansion, and the roots of the $1 \times 1$ or $2 \times 2$ minors of the matrix $\mathbf{S}$ [Eq.~(\ref{eq:3detS})] which contain $N^\phi(\phi)$ can be used to compute the appropriate coefficients. This argument can be extended by induction to the case of $N>3$ fields.

An immediate and important corollary of Eq.~\eqref{eq:detS33Expansion} is that there are no corrections to ${\cal S}$ which are linear in $\alpha_{ij}$ with $i\neq j$.
Moreover, by looking at the second and third lines of Eq.~\eqref{eq:detS33Expansion} we conclude that couplings give quadratic corrections only if $\alpha_{ij}\neq 0$ and $\alpha_{ji}\neq 0$ for $i\neq j$.
In conclusion, a perturbative coupling between nondegenerate operators $L_i$ and $L_j$ gives {\em at most quadratic corrections} to the QNM frequencies. 

\section{Degenerate spectra}
\label{app:Degenerateomega1derivation}

There is a well-known analogy between QNMs for the potential $V$ and quasibound states for the potential $-V$~\cite{Ferrari:1984zz}.  When we consider perturbations of the effective potential, following this analogy and Refs.~\cite{Leung:1999rh, Leung:1999iq}, we can compute QNM frequencies by applying quantum mechanical perturbation theory.  It is reasonable that these considerations should extend to the coupled system considered in this paper.  In this appendix we show that we can use quantum mechanical perturbation theory for a coupled system which is degenerate at zeroth order. We mostly follow standard notation from nonrelativistic quantum mechanics.

Let $H_0$ be a Hamiltonian with a degenerate eigenvalue $\omega_0$ and two corresponding eigenstates $|\omega_0,+\rangle$, $|\omega_0,-\rangle$.
A generic eigenstate with eigenvalue $\omega_0$ is given by the linear superposition 
\begin{equation}
    |\omega_0\rangle = c_1 |\omega_0\,,+\rangle + c_2 |\omega_0\,,-\rangle\,.
  \end{equation}
We denote the states by ``$\pm$'' indices rather than numerical indices because we are mainly interested in polar and axial gravitational perturbations, but our discussion below is generic.

We wish to find the spectra of the operator 
$H_0 + \delta V$,
where $\delta V$ is a small correction to the Hamiltonian.
Consider the eigenvalue problem
\begin{equation}
    (H_0 +  \delta V) |\omega\rangle = \omega |\omega\rangle\,.
    \label{eq:QMMasterPM}
\end{equation}
To first order in perturbation theory, the eigenvalues and eigenfunctions can be written as
$\omega = \omega_0 + \omega_1$
and
$|\omega\rangle = |\omega_0\rangle + |\omega_1\rangle$\,,
where
$\omega_1$ and $|\omega_1\rangle$ are first-order corrections. 
Then Eq.~\eqref{eq:QMMasterPM} becomes a relation between first-order quantities:
\begin{equation}
    ( \delta V - \omega_1 )|\omega_0\rangle = (\omega_0 - H_0) |\omega_1\rangle\,.
\label{eq:QNMExpansion}
\end{equation}
Using the relation $\langle \omega_0,\pm|(\omega_0 - H_0) = 0$ and acting to the left with $\langle \omega_0,\pm|$, one obtains
\begin{align}
    c_+   \delta V_{++} + c_-   \delta V_{+-} - \omega_1 c_+ &=0
\\
    c_+   \delta V_{-+}+ c_-  \delta V_{--} - \omega_1 c_- &=0,
\end{align}
where $ \delta V_{\pm\pm} := \langle \omega_0,\pm|\delta V|\omega_0,\pm\rangle$, and similar relations define the off-diagonal terms.
This can be written in matrix form as
\begin{align}
    \begin{pmatrix}
    \delta V_{++} - \omega_1 &  \delta V_{+-}
    \\
    \delta V_{-+} & \delta  V_{--} - \omega_1 
    \end{pmatrix}
    \begin{pmatrix}
    c_+
    \\
    c_-
    \end{pmatrix} 
    =0.
\end{align}
To have a nontrivial solution for $c_\pm$, the determinant must be zero. This yields Eq.~(\ref{eq:DegLinearE1Main}).

Fortunately, while the QNM frequency correction $\omega_1$ is nonlinear in the expectation values, the expectation values themselves are linear.
If we expand the potential as in Eq.~\eqref{eq:dV}, then the expectation values are given by
\begin{align}
  \delta V_{\pm\pm} &= \sum_{k =0}^\infty \alpha_{\pm\pm}^{(k)} \langle \omega_0,\pm |f(r)\frac{r_H^{k-2}}{r^k} |\omega_0,\pm\rangle
                      \nonumber \\
    &= \sum_{k =0}^\infty \alpha_{\pm\pm}^{(k)} \delta V^{(k)}_{\pm\pm} \,,
    \label{eq:deltaVkppDef}
\\
    \delta V_{\pm\mp} &= \sum_{k =0}^\infty \alpha_{\pm\mp}^{(k)} \langle \omega_0,\pm |f(r)\frac{r_H^{k-2}}{r^k} |\omega_0,\mp\rangle
                      \nonumber \\
    &= \sum_{k =0}^\infty \alpha_{\pm\mp}^{(k)} \delta V^{(k)}_{\pm\mp} \,.
    \label{eq:deltaVkpmDef}
\end{align}

\bibliography{QNMTaylor2_refs}
\end{document}